\documentclass[12pt]{article}
\usepackage{epsf,amssymb,amsmath,latexsym}
\usepackage{verbatim}

\usepackage{color}
 
\newcommand{\blue}{\textcolor[rgb]{0.00,0.00,0.99}} 

\usepackage{tikz}
\usepackage[utf8]{inputenc}
\usepackage{pgfplots} 
\usepackage{pgfgantt}
\usepackage{pdflscape}
\pgfplotsset{compat=newest} 
\pgfplotsset{plot coordinates/math parser=false}

\title{Oscillation theory for the density of states
\\
of high dimensional random operators}

\author{Julian Gro\ss  mann$^{1,2}$, Hermann Schulz-Baldes$^{1,3,*}$, Carlos Villegas-Blas$^{3,4}$
\\
\\
{\small $^1$ Department Mathematik, Friedrich-Alexander-Universit\"at
Erlangen-N\"urnberg, Germany}
\\
{\small $^2$ Institut f\"ur Mathematik, Technische Universit\"at Hamburg, Germany}
\\
{\small $^3$ Instituto de Matem\'aticas, UNAM, Unidad Cuernavaca, Mexico}
\\
{\small $^4$ Laboratorio Solomon Lefschetz, Unidad Mixta Internacional del CNRS, Cuernavaca, Mexico} 
\\
{\small $^*$ corresponding author, contact details: Cauerstr. 11, D-91058 Erlangen, schuba@mi.uni-erlangen.de} 
}

\date{ }

\newtheorem{theo}{Theorem}

\newtheorem{proposi}[theo]{Proposition}
\newtheorem{lemma}[theo]{Lemma}
\newtheorem{coro}[theo]{Corollary}

\newcommand{\CM}{{\mathbb C}}
\newcommand{\NM}{{\mathbb N}}
\newcommand{\RM}{{\mathbb R}}

\newcommand{\SM}{{\mathbb S}}

\newcommand{\ZM}{{\mathbb Z}}
\newcommand{\LM}{{\mathbb L}}

\newcommand{\PM}{{\mathbb P}}
\newcommand{\UM}{{\mathbb U}}
\newcommand{\Aa}{{\cal A}}
\newcommand{\Pp}{{\cal P}}

\newcommand{\EE}{{\bf E}}

\newcommand{\Tr}{\mbox{\rm Tr}}
\newcommand{\Tt}{{\cal T}}

\newcommand{\Nn}{{\cal N}}
\newcommand{\Mm}{{\cal M}}
\newcommand{\Cc}{{\cal C}}
\newcommand{\Jj}{{\cal J}}
\newcommand{\Ii}{{\cal I}}

\newcommand{\Kk}{{\cal K}}
\newcommand{\Hh}{{\cal H}}

\newcommand{\one}{\mathbf{1}}

\newcommand{\trans}{{\cal M}}
\newcommand{\ii}{{\imath}}

\begin{document}

\maketitle

\begin{abstract}
Sturm-Liouville oscillation theory is studied for Jacobi operators with block entries given by covariant operators on an infinite dimensional Hilbert space. It is shown that the integrated density of states of the Jacobi operator is approximated by the winding of the Pr\"ufer phase w.r.t. the trace per unit volume. This rotation number can be interpreted as a spectral flow in a von Neumann algebra with finite trace. 
\end{abstract}

%
%
%
%

\vspace{.2cm}

\section{Overview}
\label{sec-main}

Oscillation theory for scalar Sturm-Liouville problems and finite difference equations is a classical subject with numerous contributions dating back to the 1800's. A good historic account can be found in \cite{AHP}. The first results on matrix-valued Sturm-Liouville equations seem to be due to Bott in 1956 \cite{Bot} and for a summary of subsequent results on this matter we refer to the large bibliography in the recent book \cite{JONNF} and paper \cite{GZ}.  Bott also stressed the importance of intersection theory of Lagrangian subspaces in this context and invented an associated index theory which later on was further developed in the works of Maslov \cite{Mas}. This index can readily be read off the matrix-valued Pr\"ufer phase, as stressed in \cite{SB,SB2} which focussed on oscillation theory for matrix-valued Jacobi matrices. For operators on an infinite interval, the solutions have infinitely many oscillations, but if the coefficients of the Sturm-Liouville or finite difference equation are periodic, quasiperiodic or random, one can define an averaged rotation number and then connect it to the density of states, see \cite{PF,JM} for the one-dimensional case and \cite{SB} for matrix-valued random systems. 

\vspace{.2cm}

For Sturm-Liouville and block Jacobi operators with infinite dimensional fibers, it is still possible to define a Pr\"ufer phase as a unitary operator on an infinite dimensional Hilbert space, see below. We are not aware of any results connecting this Pr\"ufer phase to spectral properties of the initial operator, except for \cite{ASV} where a Fredholm condition allowed to focus on a single discrete eigenvalue of the Pr\"ufer phase.  In general, the spectrum of the unitary Pr\"ufer phase can be essential and spread out over the whole unit circle, so that on first sight it seems of little use for oscillation theory. However, we show below that the whole spectrum of the Pr\"ufer phase moves around the unit circle in the same direction (as a function of the energy) and therefore, under certain conditions on the fiber operators, one can define an associated spectral flow in the generalized sense of finite von Neumann algebras \cite{BCP}. To be in this framework, we consider block Jacobi matrices with operator entries taken from the C$\mbox{*}$\hbox{-}algebra of covariant operator families on $\ell^2(\ZM^d)$. This allows to study higher-dimensional discrete random Schr\"odinger operators and calculate their integrated density of states (IDOS) via the generalized spectral flow of the Pr\"ufer phases. To our best knowledge, this is an essential novel extension of oscillation theory.

\vspace{.2cm}

Let us now describe the framework and result in a more detailed manner and provide some technical insights to its proof. A Jacobi matrix of size $N\geq 3$ is a selfadjoint operator of the following tridiagonal form
\begin{equation}
\label{eq-matrix}
H_N
\;=\;
\left(
\begin{array}{ccccccc}
V_1       & T_2  &        &        &         &    \\
T_2^*      & V_2    &  T_3  &        &         &        \\
            & T_3^* & V_3    & \ddots &         &        \\
            &        & \ddots & \ddots & \ddots  &        \\
            &        &        & \ddots & V_{N-1} & T_N   \\
    &        &        &        & T_N^*  & V_N
\end{array}
\right)
\;.
\end{equation}
In the classical case, the entries $(V_n)_{n=1,\ldots,N}$ and $(T_n)_{n=2,\ldots,N}$ are real numbers and it is a well-known how to calculate the spectrum of $H_N$ by Sturm-Liouville oscillation theory \cite{AHP}. For selfadjoint $V_n$ and invertible $T_n$ matrices of same size a corresponding result is spelled out \cite{SB,SB2}, see also \cite{DK,KHZ} and \cite{JONNF}. Here we will be interested in the situation where the entries $V_n=V_n^*$ and $T_n$ are  covariant and local operator families on $\ell^2(\ZM^d)$ which are representations of elements of a crossed product C$\mbox{*}$\hbox{-}algebra $\Aa=C(\Omega)\rtimes \ZM^d$ associated to a compact dynamical system $(\Omega,\tau,\ZM^d)$. Here $\Omega$ is a compact space  (of disorder configurations) equipped with a continuous action $\tau=(\tau_1,\ldots,\tau_d)$ of the group $\ZM^d$ \cite{Bel,PF}. More concretely, such a covariant family $A=(A_\omega)_{\omega\in\Omega}$ consists of local operators $A_\omega$ on $\ell^2(\ZM^d)$ which are strongly continuous in $\omega\in\Omega$. The locality property means that matrix elements $\langle n|A_\omega|m\rangle$ decay sufficiently fast with the distance $|n-m|$ between points $n,m\in\ZM^d$. Moreover, the covariance property w.r.t. the right shifts  $S_1,\ldots,S_d$ on $\ell^2(\ZM^d)$ holds:
\begin{equation}
\label{eq-cov}
S_j\, A_\omega\,S_j^*
\;=\;
A_{\tau_j\omega}
\;,
\qquad
j=1,\ldots,d
\;.
\end{equation}
Covariant and local operator families are also called homogeneous \cite{Bel} and metrically transitive \cite{PF}. If $\PM$ is an invariant and ergodic probability measure on $\Omega$, one obtains a normalized and finite trace on $\Aa$ by setting
\begin{equation}
\label{eq-trace}
\Tt(A)
\;=\;
\EE\;\langle 0|A_\omega|0\rangle
\;,
\end{equation}
where $\EE$ denotes the average w.r.t. $\PM$. For more details on $\Aa$ and $\Tt$ the reader is referred to \cite{Bel} and Section~\ref{sec-covariant} below. Each of the entries $(T_n,V_n)$ is then drawn independently and identically with $\PM$. If $\ZM^0$ is by convention identified with one point, the case $d=0$ corresponds to a scalar Jacobi matrix since the covariance condition is then empty. For sake of concreteness, we will stick to the framework just described, but note that one can consider the covariance relation w.r.t. magnetic translations (at the expense of working with a twisted crossed product algebra $\Aa$)  and can choose the space $\Omega$ as well as the probability $\PM$  to be dependent on $n$ (at least in the first part of Theorem~\ref{theo-osci} below). Apart from the covariance, we will suppose that the coefficient operators satisfy the following:

\vspace{.3cm}

\noindent {\bf Standing assumption:} $\Lambda =\sup_{n\geq 1}\{\|T_n\|,\|T^{-1}_n\|,\|V_n\|\}<\infty$ 

\vspace{.3cm}

\noindent Resuming, $H_N$ is a discrete Schr\"odinger operator on the strip Hilbert space $\ell^2(\{1,\ldots,N\}\times\ZM^d)$ with matrix elements of range $1$ in the finite direction $\{1,\ldots,N\}$ and a covariance relation in the fiber $\ZM^d$. A typical example where all the above holds is the standard Anderson model on the strip $\{1,\ldots,N\}\times\ZM^d$. There $H_N$ is of the form \eqref{eq-matrix} with $T_n=\one$ the identity on $\ell^2(\ZM^d)$ and 
$$
V_n\;=\;\sum_{j=1}^d(S_j+S_j^*)\;+\;
\sum_{m\in\ZM^d} v_{n,m}\,|m\rangle\langle m|\;:\;\ell^2(\ZM^d)\to\ell^2(\ZM^d)
$$
where $\omega=(v_{n,m})_{n\in\ZM,m\in\ZM^d}\in\Omega$ are i.i.d. real random variables distributed by a compactly supported probability measure.

\vspace{.2cm}

As a matrix-valued covariant operator family, $H_N$ has a well-defined IDOS given by
$$
\Nn_N(E)
\;=\;
\frac{1}{N}\;\Tr_N\otimes \Tt\big(\chi(H_N\leq E)\big) 
\;,
$$
where $\chi$ is the characteristic function and $\Tr_N$ the trace over $\CM^N$. Note that $E\in\RM\mapsto\Nn_N(E)$ is increasing with limits $\Nn_N(-\infty)=0$ and $\Nn_N(\infty)=1$, hence it specifies a probability measure called the density of states (DOS) of the covariant operator family $H_N$ on the strip Hilbert space $\ell^2(\{1,\ldots,N\}\times\ZM^d)$. In the limit $N\to\infty$, one obtains the IDOS of the $(d+1)$-dimensional covariant  Hamiltonian on $\ell^2(\ZM^{d+1})$:
$$
\Nn(E)
\;=\;
\lim_{N\to\infty}\Nn_N(E)
\;.
$$
The main aim of this paper is to calculate the integrated density of states $\Nn_N(E)$ by means of Sturm-Liouville type oscillation theory, namely to approximate it by an average rotation number.  

\vspace{.2cm}

For that purpose, let us begin by introducing the main objects of oscillation theory, that is, transfer operators and Pr\"ufer phases. The transfer operators are $2\times 2$ matrices with entries in $\Aa$ given by
\begin{equation}
\label{eq-transfer}
\trans_n^E
\;=\;
\left(
\begin{array}{cc}
(E\,{\bf 1}\,-\,V_n)\,T_n^{-1} & - T_n^* \\
T_n^{-1} & {\bf 0}
\end{array}
\right)
\;,
\qquad
n=1,\ldots,N
\;,
\end{equation}
where $T_1=\one$. Then the Pr\"ufer phase is defined by
\begin{equation}
\label{eq-Pruefer}
U^E_N\;=\;
\left[\binom{\one}{\imath\,\one}^*  \trans_N^E\cdots\trans_1^E \binom{\one}{0}\right]\,
\left[\binom{\one}{-\imath\,\one}^*\trans_N^E\cdots\trans_1^E\binom{\one}{0}\right]^{-1}\;.
\end{equation}
By using Krein space techniques, it will be shown in Section~\ref{sec-prelim} below that $U^E_N$ is a well-defined and unitary operator in $\Aa$. In particular, for $d=0$ it is simply a number on the unit circle. The following monotonicity is crucial for oscillation theory:

\begin{proposi}
\label{prop-derivbound}
For $E\in\RM$ and $N\geq 2$, the phase velocity operator
$$
S^E_N
\;=\;
\frac{1}{\imath\,N}\,({U}^E_N)^*\,\partial_E\,{U}^E_N
\;\in\;\Aa
$$
is positive semi-definite and 
$$
\int^\infty_{-\infty} dE\;\Tt(S^E_N)\;=\;2\pi\;.
$$
\end{proposi}

In the classical case $d=0$, this means that the Pr\"ufer phase always rotates in the positive sense $N$ times around the unit circle as a function of energy $E$. Each time it passes by $-1$ the matrix $H_N$ has an eigenvalue. In the situation of a finite dimensional fiber (so still $d=0$, but with finite dimensional matrices $T_n$ and $V_n$), the Pr\"ufer phase is a finite dimensional unitary matrix and with eigenvalues all rotating in the positive sense on the unit circle, and again each time that one passes by $-1$  there is an eigenvalue of $H_N$ \cite{SB}. This follows from intersection theory for pairs of Lagrangian subspaces (the solution at $N$, and the boundary condition at $N$ respectively), which is calculated by the Bott-Maslov index. In the present situation of infinite dimensional fibers, however, the spectrum of ${U}^E_N$ is essential and in interesting examples actually fills the whole unit circle. We do not attempt to define intersection theory in this situation. On the other hand, it does make sense to consider the spectral flow around the unit circle w.r.t. the trace $\Tt$. That this spectral flow indeed approximates the IDOS of $H_N$ is the main result of the paper:

\begin{theo} 
\label{theo-osci}
Let $H_N$ be a Jacobi operator with local, covariant and i.i.d. matrix entries satisfying the standing assumption. We suppose that also finite volume restrictions of $T_n$ and $V_n$ satisfy the standing assumption, see Section~\ref{sec-FiniteApp} below for details. Then there exists a constant $C$ such that uniformly in $E$
\begin{equation}
\label{eq-MainEst}
\left|
\,\Nn_N(E)\,-\,
\frac{1}{2\pi}\,\int^E_{-\infty} de\;\Tt(S^e_N)\,
\right|
\;\leq\;
\frac{C}{N}
\;.
\end{equation}
In particular, the $(d+1)$-dimensional IDOS satisfies
$$
\Nn(E)
\;=\;
\lim_{N\to\infty}
\;\frac{1}{2\pi}\,\int^E_{-\infty} de\;\Tt(S^e_N)
\;.
$$
\end{theo}

In the case $d=0$, the quantity $\int^E_{-\infty} de\;\Tt(S^e_N)$ is called rotation number (up to a factor $N^{-1}$) and then Theorem~\ref{theo-osci} is a well-known result \cite{PF,AHP}. For general $d$, it is the integrated phase velocity density w.r.t. $\Tt$ and we also refer to it as a rotation number. It is a type II$_1$ spectral flow in the finite von Neumann algebra $L^{\infty}(\Aa,\Tt)$ as defined in  \cite[Section~5.1]{BCP}. The core of the proof of Theorem~\ref{theo-osci} is a result from \cite{SB}, see Theorem~\ref{theo-MaslovEst} in Section~\ref{sec-FiniteApp}, which states that the estimate \eqref{eq-MainEst} holds for Jacobi matrices with finite dimensional matrix entries, {\it independently} of the size of these matrix entries. This allows to reduce the proof of Theorem~\ref{theo-osci} to a control of finite size approximations of the IDOS and the rotation number. While this is standard for the IDOS \cite{PF}, the approximation of the rotation number requires precise high energy estimates for the Pr\"ufer phases and its derivatives. This part of the proof, given in Section~\ref{sec-Pruefer}, is unfortunately rather technical, but unavoidable in our opinion. We are not aware of prior works on such asymptotics in the scalar case, but suspect that they may not be novel in this case.

\vspace{.3cm}

\noindent {\bf Acknowledgement:}  H.~S.-B. thanks Alan Carey for discussions at a preliminary early stage of this project. This work was partially supported by the PREI-UNAM, PAPIIT-UNAM-IN104015 as well as the DFG.

\vspace{.2cm}

\section{Preliminaries}
\label{sec-prelim}

\subsection{Operators on Krein space}
\label{sec-Krein}

Let $\Hh$ be a Hilbert space with a scalar product denoted by $\langle v|w\rangle$ for $v,w\in\Hh$. The doubled Hilbert space $\Kk=\Hh\oplus\Hh$ becomes a Krein space when equipped with either a selfadjoint unitary $\Jj$ or a skew-adjoint unitary $\Ii$ which can be chosen to be
\begin{equation}
\label{eq-JIdef}
\Jj\;=\;
\begin{pmatrix}
\one & 0 \\
0 & -\one
\end{pmatrix}
\;,
\qquad
\Ii\;=\;
\begin{pmatrix}
0 & -\one \\
\one & 0
\end{pmatrix}
\;.
\end{equation}
These two operators induce two quadratic forms $v\in\Kk \mapsto \langle v|\Jj|v\rangle$ and $v\in\Kk \mapsto \langle v|\Ii|v\rangle$ on $\Kk$. Of course, these forms and the two operators $\Jj$ and $\Ii$ are related, namely 
\begin{equation}
\label{eq-IJlink}
\imath\; \Ii\;=\;\Cc^*\, \Jj\,\Cc
\;,
\end{equation}
where $\Cc$ is the Cayley transform given by
\begin{equation}
\label{eq-Cayley}
{\Cc}
\;=\;
\frac{1}{\sqrt{2}}\;
\left(
\begin{array}{cc}
\one & -\imath\,\one \\
\one & \imath\,\one
\end{array}
\right)
\;.
\end{equation}
The quadratic forms are conserved by the so-called $\Jj$-unitary and $\Ii$-unitary operators on $\Kk$. These are bounded linear operators $\Nn$ and $\Mm$ on $\Kk$ which satisfy $\Nn^*\Jj\Nn=\Jj$ and $\Mm^*\Ii\Mm=\Ii$ respectively. These operators form two subgroups of the general linear group $\mbox{\rm GL}(\Kk)$ on $\Kk$:
$$
\UM(\Kk,\Jj)
\;=\;
\left\{ \Nn\in\mbox{\rm GL}(\Kk)\;:
\;\Nn^*\Jj\Nn=\Jj\right\}
\;,
\qquad
\UM(\Kk,\Ii)
\;=\;
\left\{ \trans\in\mbox{\rm GL}(\Kk)\;:
\;\trans^*\Ii\trans=\Ii\right\}
\;.
$$
In particular, note that when $\Nn\in \UM(\Kk,\Jj)$, then also the inverse $\Nn^{-1}$ is in $\UM(\Kk,\Jj)$. Taking the inverse of the relation $\Nn^*\Jj\Nn=\Jj$ also shows that $\Nn^*$ is in $\UM(\Kk,\Jj)$. It follows from \eqref{eq-IJlink} that
$$
\Cc^*\,\UM(\Kk,\Jj)\,\Cc\;=\;\UM(\Kk,\Ii)
\;.
$$
The groups can be written out more explicitly using linear operators $A,B,C,D$ on $\Hh$, for example:
\begin{align}
\label{eq-ULL}
& 
\UM(\Kk,\Jj) 
\;=\;
\left\{
\,\left(
\begin{array}{cc}
A & B \\  C & D
\end{array}
\right)
\in\mbox{\rm GL}(\Kk)
\;
:
\;
A^*A-C^*C=\one\;,\;
D^*D-B^*B=\one\;,\;
A^*B=C^*D\;
\right\}
\;.
\end{align}
Relevant properties in the following will be $D^*D=\one+B^*B\geq \one$ and thus $\|D^{-1}\|\leq 1$. As also $\Nn^*\in \UM(\Kk,\Jj)$, one further has $DD^*=\one+CC^*$ so that $D^{-1}C(D^{-1}C)^*=\one-D^{-1}(D^{-1})^*< \one$. In particular, $\|D^{-1}C\|< 1$. We need improved bounds for transfer operators of the form \eqref{eq-transfer}.

\begin{proposi}
\label{prop-Transfer}
Let $V=V^*$  be a selfadjoint bounded operator on $\Hh$ and $T$ an invertible bounded operator on $\Hh$. For any $E\in\RM$, one then has
$$
\Mm^E
\;=\;
\begin{pmatrix}
(E-V)T^{-1} & -T^* \\ T^{-1} & 0
\end{pmatrix}
\;\in\;
\UM(\Kk,\Ii)
\;.
$$
Furthermore, 
\begin{equation}
\label{eq-CayleyTransfer}
\Cc\,\Mm^E\,\Cc^*
\;=\;
\frac{1}{2}
\begin{pmatrix}
(E-V)T^{-1} -\imath(T^*+ T^{-1}) 
& (E-V)T^{-1} +\imath(T^*- T^{-1}) 
\\
(E-V)T^{-1} -\imath(T^*- T^{-1}) 
&
(E-V)T^{-1} +\imath(T^*+ T^{-1}) 
\end{pmatrix}
\;\in\;
\UM(\Kk,\Jj)
\;.
\end{equation}
In the representation \eqref{eq-ULL}, the entries $C$ and $D$ of $\Cc\Mm^E\Cc^*$ satisfy
$$
\|D^{-1}C\|^2
\;\leq\; 
1\,-\,4\,\Lambda^{-2}(|E|\,+\,\Lambda\,+\, 2 )^{-2}
\;,
$$
where $\Lambda=\max\{\|V\|,\|T\|,\|T^{-1}\|\}$.
\end{proposi}

\noindent {\bf Proof.} The first two claims are short calculations, so let us focus on the estimate. We will verify $D^{-1}(D^{-1})^*\geq\mu^2\,\one$ for $\mu=2\Lambda^{-1}(|E|\,+\,\Lambda\,+\, 2 )^{-1}$. Due to $D^{-1}C(D^{-1}C)^*=\one-D^{-1}(D^{-1})^*$, this will imply the result. But $D^{-1}(D^{-1})^*\geq\mu^2\,\one$  is equivalent to $DD^*\leq \mu^{-2}\,\one$ and thus follows from $\|D\|\leq\mu^{-1}$. Thus
$$
\|D\|\;=\;
\frac{1}{2}\;
\|
(E-V)T^{-1} +\imath(T^*+ T^{-1})\|
\;\leq\;
\frac{1}{2}\;
(|E|+\Lambda)\Lambda\,+\,\Lambda
$$
concludes the proof.
\hfill $\Box$

\begin{proposi}
\label{prop-Transfer2}
Let $\Mm_n^E$ be defined as in \eqref{eq-transfer} and suppose that the standing assumption $\Lambda =\sup_{n\geq 1}\{\|T_n\|,\|T^{-1}_n\|,\|V_n\|\}<\infty$ holds. Define $A,B,C,D$ by
\begin{equation}
\label{eq-MN1Coef}
\Cc\,\Mm_N^E\cdots \Mm_1^E\,\Cc^*
\;=\;
\begin{pmatrix}
A & B \\ C & D
\end{pmatrix}
\;.
\end{equation}
Then there exists a positive constant $\mu_N=\mu_N(\Lambda,E)>0$ such that
$$
\|D^{-1}C\|^2
\;\leq\; 
1\,-\,\mu_N^2
\;.
$$
\end{proposi}

\noindent {\bf Proof.} For $N=1$ this is precisely Proposition~\ref{prop-Transfer} with $\mu_1=\Lambda^{-1}(|E|\,+\,\Lambda\,+\, 2 )^{-1}$. For general $N$, let us note that $D$ is a polynomial in $E$ of order $N$ with coefficients given by the $V_n$, $T_n$ and $(T_n)^{-1}$ which are all estimated by $\Lambda$. Thus $\|D\|\leq c_N|E|^N\Lambda^N$ for some constant $c_N$ and $E$ sufficiently large. Now the argument of the proof of Proposition~\ref{prop-Transfer} leads to the stated bound for large $E$. For small $E$, one has $\|D^{-1}C\|< 1$ as pointed before Proposition~\ref{prop-Transfer}. Hence by a compactness argument the bound on $\|D^{-1}C\|$ holds uniformly in energy.
\hfill $\Box$

\subsection{Lagrangian subspaces}
\label{sec-Lagrange}

This section recollects a few basic facts needed below, which can all be found, {\it e.g.}, in  \cite{SB3}. A subspace of a Krein space $(\Kk,\Jj)$ is called $\Jj$-Lagrangian if $\Jj$ viewed as a quadratic form $v \mapsto \langle v|\Jj|v\rangle$ vanishes on it and it is maximal with this property. Let $\LM(\Kk,\Jj)$ denote the Lagrangian Grassmannian, namely the set of all Lagrangian subspaces of $(\Kk,\Jj)$. A $\Jj$-Lagrangian frame on $\Kk$ is a linear operator $\Psi:\Hh\to\Kk$ such that $\Psi^*\Psi=\one$, $\Psi^*\Jj\Psi=0$ and the range of $\Psi$ is a Lagrangian subspace. Note that, if $\Psi$ is a Lagrangian frame, then so is $\Psi U$ where $U$ is a unitary operator on $\Hh$ and all equivalence classes $[\Psi]$ of Lagrangian frames w.r.t. this right action by the unitary group $\UM(\Hh)$ on $\Hh$ constitute the Lagrangian Grassmannian $\LM(\Kk,\Jj)$. All these notions transpose directly to the Krein space $(\Kk,\Ii)$. 

\begin{proposi}
\label{prop-diffeo} 
The Lagrangian Grassmannians are bijectively mapped onto the unitary operators $\UM(\Hh)$ on $\Hh$ by the stereographic projections $\Pi_\Jj:\LM(\Kk,\Jj)\to \UM(\Hh)$ and $\Pi_\Ii:\LM(\Kk,\Ii)\to \UM(\Hh)$ defined by
$$
{\Pi}_\Jj\bigl([{\Psi}]\bigr)
\;=\;
\left[\binom{\one}{0}^*{\Psi}\right]\,
\left[\binom{0}{\one}^*{\Psi}\right]^{-1}
\;,
\qquad
{\Pi}_\Ii\bigl([{\Phi}]\bigr)
\;=\;
\left[\binom{\one}{\imath\,\one}^*{\Phi}\right]\,
\left[\binom{\one}{-\imath\,\one}^*{\Phi}\right]^{-1}\;,
$$
where $\Psi$ and $\Phi$ are $\Jj$-Lagrangian and $\Ii$-Lagrangian frames, respectively.
\end{proposi}

Let us point out that $\Cc^* \Psi$ is an $\Ii$-Lagrangian frame if $\Psi$ is a $\Jj$-Lagrangian frame and that ${\Pi}_\Jj\bigl([{\Psi}]\bigr)={\Pi}_\Ii\bigl([\Cc^*{\Psi}]\bigr)$. Also let us note that the inverse of $\Pi_\Ii$ is given by
\begin{equation}
\label{eq-Piinv}
\Pi^{-1}_\Ii(U)
\;=\;
\left[
\frac{1}{2}
\left(
\begin{array}{c} (U+\one) \\  \imath\,(U-\one)
\end{array}\right)
\right]
\;.
\end{equation}

The group $\UM(\Kk,\Ii)$ naturally acts on $\LM(\Kk,\Ii)$. Explicitly, $(\trans,[\Phi])\in \UM(\Kk,\Ii)\times\LM(\Kk,\Ii)\mapsto [\trans\Phi|\trans\Phi|^{\blue{-1}}]\in\LM(\Kk,\Ii)$.  Under the stereographic projection $\Pi_\Ii$ this action becomes the action of the group  $\UM(\Kk,\Jj) $ via operator M\"obius transformation on the unitary group:
\begin{equation}
\label{eq-Moeb}
\Pi_\Ii\bigl([\trans\Phi]\bigr)
\;=\;
\Cc\trans\Cc^*\cdot\Pi_\Ii\bigl([\Phi]\bigr)
\;,
\end{equation}
where the dot denotes
$$
\left(
\begin{array}{cc}
A & B \\
C & D
\end{array}
\right)
\cdot U
\;=\;
(A\,U+B)(C\,U+D)^{-1}
\;.
$$

\subsection{Covariant operators}
\label{sec-covariant}

Let us briefly recall from \cite{Bel} the definition of the (reduced) crossed product algebra $\Aa$. One first starts with the algebra $\Aa_0$ of strongly continuous  families $A=(A_\omega)_{\omega\in\Omega}$ of bounded operators on $\ell^2(\mathbb{Z}^d)$ which satisfy the covariance relation \eqref{eq-cov} and are of finite range, notably there exists an $R$ such that $\langle n|A_\omega|m\rangle=0$ for $|n-m|>R$ for all $\omega$. On $\Aa_0$ there is a C$^*$-norm
$$
\|A\|
\;=\;
\sup_{\omega\in\Omega}
\|A_\omega\|
\;,
$$
and $\Aa$ is the completion of $\Aa_0$ under this norm. In the following, matrices with entries from $\Aa$ will simply be called covariant and local. Now \eqref{eq-trace} defines a finite trace on $\Aa$ which, by Birkhoff's ergodic theorem, is $\PM$-almost surely equal to the trace per unit volume
$$
\Tt(A)
\;=\;
\lim_{L\to\infty}\;\frac{1}{(2L+1)^d}\;\sum_{|n|_\infty\leq L}\,\langle n|A_\omega|n\rangle
\;,
$$
where $|n|_\infty=\max\{|n_1|,\ldots,|n_d|\}$. For the study of the covariant and local Jacobi operators $H_N$ described in Section~\ref{sec-main}, it will be necessary to consider covariant and local families $\Phi=(\Phi_\omega)_{\omega\in\Omega}$ of Lagrangian frames in the Krein space $\Kk=\ell^2(\ZM^d)\otimes\CM^2$ equipped  with either $\Jj$ or $\Ii$, notably 
\begin{equation}
\label{eq-CovLag}
S_j\oplus S_j\;\Phi_\omega\,S_j^*
\;=\;
\Phi_{\tau_j\omega}
\;,
\qquad
j=1,\ldots,d
\;.
\end{equation}
The ranges of these frames specifies a covariant and local family of Lagrangian subspaces on the Krein space (on which the group $\ZM^d$ acts by $S_j\oplus S_j$). 

\begin{coro}
\label{coro-diffeo} 
The stereographic projections $\Pi_\Jj$ and $\Pi_\Ii$ identify the set of covariant and local Lagrangian subspaces with the covariant unitary operators, namely the unitaries in $\Aa$.
\end{coro}

As in the theory of the Bott-Maslov index \cite{Bot}, it is now natural to study the intersection of two covariant families of Lagrangian subspaces. This intersection will again be a covariant family of subspaces and one can measure its (non-integer) dimension using $\Tt\oplus\Tt(\Pp)$ where $\Pp$ is the covariant family of orthogonal projections onto the intersection. In interesting (and actually typical) cases, $\Tt\oplus\Tt(\Pp)$ vanishes. Hence intersection theory is of little interest in this context. On the other hand, the flow of intersection density can be measured by a spectral flow and this is what is ultimately used in Theorem~\ref{theo-osci}.

\vspace{.2cm}

Just as it is possible to consider covariant and local families of Lagrangian plane in \eqref{eq-CovLag}, one can also consider covariant families in the group $\UM(\Kk,\Jj)$. This merely means that the matrix entries of these operators, {\it e.g.} in \eqref{eq-ULL}, are taken from $\Aa$. This set is a subgroup of $\UM(\Kk,\Jj)$ and it acts on the covariant and local $\Jj$-Lagrangian planes, and under the stereographic projection this leads to an action of the covariant and local $\Jj$-unitaries on the unitary group in $\Aa$. The same statements hold in the Krein space $(\Kk,\Ii)$.

\vspace{.2cm}

\section{Analysis of Pr\"ufer phases}
\label{sec-Pruefer}

\subsection{Eigenfunctions and transfer matrices}
\label{sec-transfer}

As for a one-dimensional Jacobi matrix, it is useful to rewrite the formal Schr\"odinger equation 
\begin{equation}
\label{eq-Schroedinger}
H_N\phi
\;=\;
E\,\phi\;, 
\end{equation}
for a real energy $E\in\RM$ in terms of the transfer operators $\trans_n^E$ defined by \eqref{eq-transfer} and analyzed in Proposition~\ref{prop-Transfer}. Here we view \eqref{eq-Schroedinger} as an operator equation for the components of $\phi=(\phi_1,\ldots,\phi_N)$ which are all operators on $\Hh$. The  Schr\"odinger equation (\ref{eq-Schroedinger}) is satisfied if and only if
\begin{equation}
\label{eq-transferid}
\left(
\begin{array}{c}
T_{n+1}\phi_{n+1} \\
\phi_n
\end{array}
\right)
\;=\;
\trans^E_n\,
\left(
\begin{array}{c}
T_{n}\phi_{n} \\
\phi_{n-1}
\end{array}
\right)
\;,
\qquad \text{for all }
n=1,\ldots,N
\;,
\end{equation}
and the boundary conditions hold $\phi_{N+1}=\phi_0=0$. Here \eqref{eq-transferid} will be used to generate formal solutions
$$
{\Phi}^E_{n}
\;=\;
{\trans}^E_n
\;{\Phi}^E_{n-1}
\;,
\qquad
{\Phi}^E_0
\;=\;
\binom{\one}{0}
\;.
$$
As the range of ${\Phi}^E_0$ is $\Ii$-Lagrangian and ${\trans}^E_n$ is $\Ii$-unitary, one knows that all ${\Phi}^E_n$ are $\Ii$-Lagrangian. Moreover,  ${\Phi}^E_0$ and thus also ${\Phi}^E_n $ are covariant and local.  The stereographic projections of $\Phi^E_N$ given by the unitaries
$$
U^E_N
\;=\;
\Pi_{\Ii} ([\Phi^E_{N}])
\;,
$$
coincide with \eqref{eq-Pruefer} and are, as above, called the associated Pr\"ufer phases.  They are also covariant and local families of unitaries, and hence can be seen as elements of $\Aa$. Setting $\trans^E(N,1)=\trans^E_N\cdots\trans^E_1$ and using $\Pi_{\Ii}([\binom{\one}{0}]=\one$ as well as \eqref{eq-Moeb}, one has
\begin{equation} 
\label{eq-MoebRep}
U^E_N
\;=\;
\Cc\trans^E(N,1)\Cc^*\cdot \one\;.
\end{equation}

\subsection{Estimates on Pr\"ufer phases}
\label{sec-EstPruefer}

\begin{proposi}
\label{prop-PrueferBound}
For any finite $E_0>0$, $j\in\NM$ and $N\in\NM$ there exists a constant $K=K(E_0,j,N)$ such that the Pr\"ufer phases satisfy
$$
\sup_{E\in[-E_0,E_0]}\;
\|\partial^j_E U^E_N\|
\;\leq\;
K\;.
$$
\end{proposi}

\noindent {\bf Proof.}
Let $A,B,C,D$ be the matrix entries of $\Cc\trans^E(N,1)\Cc^*$ as given in \eqref{eq-MN1Coef}. Each of them is a polynomial of degree at most $N$ in $E$. Moreover, due to \eqref{eq-MoebRep}
$$
U^E_N
\;=\;
(A+B)(C+D)^{-1}
\;=\;
(A+B)(D^{-1}C+\one)^{-1}D^{-1}
\;.
$$
Hence deriving
$$
\partial_E\,
U^E_N
\;=\;
\Big(
\big(\partial_E A+\partial_E B\big)
\,-\,
U^E_N
\big(\partial_E C+\partial_E D\big)
\Big)
\big(D^{-1}C+\one\big)^{-1}D^{-1}
\;.
$$
Because the coefficients are polynomials in $E$, they are bounded on every compact set. Furthermore, the first inverse is bounded by Proposition~\ref{prop-Transfer2} for $E$ in the compact set $[-E_0,E_0]$, and the bound $\|D^{-1}\|\leq 1$ holds in general, see \eqref{eq-ULL}. Deriving several times and using the same bounds provides the general estimate. 
\hfill $\Box$

\subsection{Monotonicity of Pr\"ufer phases}
\label{sec-mono}

In this section, we verify Proposition~\ref{prop-derivbound} and hence the key monotonicity property which the Pr\"ufer phase has in the energy parameter.

\vspace{.2cm}

\noindent {\bf Proof} of Proposition~\ref{prop-derivbound}:  Let us introduce
$\phi^E_\pm=(\,\one\;\pm\imath\one\,)\;{\Phi}^E_{N}$. These are
invertible covariant operators (as in  Proposition~\ref{prop-diffeo}) and one has
${U}^E_N=\phi_-^E(\phi_+^E)^{-1}=((\phi_-^E)^{-1})^*(\phi_+^E)^*$. Now
$$
({U}^E_N)^*\,\partial_E\,{U}^E_N
\;=\;
((\phi_+^E)^{-1})^*
\Bigl[\,
(\phi_-^E)^*\partial_E\phi_-^E \,-\,
(\phi_+^E)^*\partial_E\phi_+^E
\,\Bigr]
(\phi_+^E)^{-1}
\;.
$$
Thus it is sufficient to verify positive definiteness of
$$
\frac{1}{\imath}\;
\Bigl[\,
(\phi_-^E)^*\partial_E\phi_-^E \,-\,
(\phi_+^E)^*\partial_E\phi_+^E
\,\Bigr]
\;=\;
2\;
({\Phi}^E_{N})^*\,\Ii\,\partial_E {\Phi}^E_{N}
\;.
$$
From the product rule follows that
$$
\partial_E {\Phi}^E_{N}
\;=\;
\sum_{n=1}^N
\;
{\trans}^E_N\cdots {\trans}^E_{n+1}
\,
\left(\partial_E {\trans}^E_n\right)
\;
{\trans}^E_{n-1} \cdots {\trans}^E_1
\,{\Phi}_0^E
\;.
$$
This implies that
$$
({\Phi}^E_{N})^*\,\Ii\,\partial_E {\Phi}^E_{N}
\;=\;
\sum_{n=1}^N
\;({\Phi}_0^E)^*\,
\left(
{\trans}^E_{n-1}\cdots {\trans}^E_1
\right)^*
\,
\bigl({\trans}^E_n\bigr)^*\,\Ii\,
\bigl(\partial_E {\trans}^E_n\bigr)
\;
\left(
{\trans}^E_{n-1}\cdots {\trans}^E_1
\right)
\,{\Phi}_0^E
\;.
$$
As one checks that
$$
\bigl({\trans}^E_n\bigr)^*\,\Ii\,
\bigl(\partial_E {\trans}^E_n\bigr)
\;=\;
\left(
\begin{array}{cc}
 (T_nT_n^*)^{-1} & {0} \\
0 &  0  
\end{array}
\right)
\;,
$$
and thus
$$
({\Phi}^E_N)^*\,\Ii\,\partial_E {\Phi}^E_N
\;=\;
\sum_{n=1}^N
\;
({\Phi}^E_0)^*
\left(
{\trans}^E_{n-1}\cdots {\trans}^E_1
\right)^*
\,
\left(
\begin{array}{cc}
(T_nT_n^*)^{-1} & {0}  \\
0 &  0
\end{array}
\right)
\;
\left(
{\trans}^E_{n-1}\cdots {\trans}^E_1
\right)
{\Phi}^E_0
\;.
$$
Clearly each of the summands is positive semi-definite. In order to prove a strict lower bound, it is sufficient that the first two
terms $n=1,2$ give a strictly positive contribution. Hence 
let us verify that
$$
\left(
\begin{array}{cc}
(T_1T_1^*)^{-1} & { 0} \\
{0} & { 0}
\end{array}
\right)
\;+\;
\bigl(\trans^E_1\bigr)^*\;
\left(
\begin{array}{cc}
(T_2T_2^*)^{-1} & {0} \\
{0} & {0}
\end{array}
\right)
\;\trans^E_1
\;>\;
0
\;.
$$
As $(T_n T_n^*)^{-1}\geq \Lambda^{-2}\,\one$ for all $n$, this positivity is equivalent to
$$
\left(
\begin{array}{cc}
\one & { 0} \\
{0} & { 0}
\end{array}
\right)
\;+\;
\bigl(\trans^E_1\bigr)^*\;
\left(
\begin{array}{cc}
\one & {0} \\
{0} & {0}
\end{array}
\right)
\;\trans^E_1
\;>\;
0
\;.
$$
Using the notation $A^E=(E-V_1)T_1^{-1}$ which in norm is bounded above by a constant (depending on $E$), one thus just has to note the invertibility
$$
\begin{pmatrix}
\one + (A^E)^*A^E & -(T_1A^E)^* \\ -T_1A^E & T_1T_1^*
\end{pmatrix}
\;= \;
\begin{pmatrix}
\one & -(A^E)^* \\ 0 & T_1
\end{pmatrix}
\begin{pmatrix}
\one & 0 \\ -A^E & T_1^\ast
\end{pmatrix}
\;.
$$
A proof of the integral identity can be shown as in Proposition~7 in \cite{SB}, or by applying that result directly to the finite volume approximations and invoking the arguments of Section~\ref{sec-Proof}.
\hfill $\Box$

\subsection{Asymptotics  of Pr\"ufer phases}
\label{sec-Asymp}

\begin{proposi}
\label{prop-asymptotics}
There are constants $C$ and $E_0$ such that for all $N\geq 1$ and $|E| > E_0$ 
$$
\|U_N^E - (\one\,-\,2\,\imath\,E^{-1})\|\;\leq\;\frac{C\,\Lambda^2}{E^2}
\;.
$$
In particular,
$$
\lim_{E\to\pm\infty}\;{U}^{E}_N\;=\;\one\;.
$$
\end{proposi}

\noindent {\bf Proof.}  We focus on the case $E>0$. Let us set
\begin{equation}
\label{eq-RUrel}
R^E_N\;=\;U^E_N\,-\,(\one\,-\,2\,\imath\,E^{-1})
\;,
\end{equation}
and  first derive an iterative equation for this operator, that is, express it in terms of $R_{N-1}^E$. This will be obtained from \eqref{eq-MoebRep}, namely $U^E_N= \Cc\trans^E_N\Cc^*\cdot U^E_{N-1}$, which becomes explicitly:
\begin{align*}
U^E_N
&\,=\,
\begin{pmatrix}
(E-V_N)T_N^{-1}   -\ii T_N^\ast - \ii T_N^{-1} & (E-V_N)T_N^{-1} + \ii T_N^\ast  -\ii T_N^{-1}
\\
(E-V_N)T_N^{-1} - \ii T_N^\ast  + \ii T_N^{-1} & (E-V_N)T_N^{-1}  + \ii T_N^\ast + \ii T_N^{-1}
\end{pmatrix}
\cdot U^E_{N-1}
\\
&\,=\,
\begin{pmatrix}
[\one - w_1(N,E)] T_N^{-1} & [\one - w_2(N,E)] T_N^{-1}
\\
[\one - w_3(N,E)] T_N^{-1} & [\one - w_4(N,E)] T_N^{-1}
\end{pmatrix}
\cdot U^E_{N-1}
\;,
\end{align*}
where we set
$$
w_1(N,E) \;=\; 
E^{-1} (V_N   +  \ii \,T_N^\ast T_N + \ii )
\;,
$$
and similarly $w_2(N,E), w_3(N,E), w_4(N,E)$ with other signs. To shorten notations, we will from now on drop the arguments of $w_j$, and also the index $N$ on $T_N$ and $V_N$. As $\Lambda\geq 1$, $\|V\| \leq \Lambda$ and $\|T\| \leq \Lambda$, the operators $w_j$ satisfy
\begin{equation}
\label{eq-wEst}
\|w_j\|\;\leq\;E^{-1}(\Lambda+\Lambda^2+1)\;\leq\;3\,E^{-1}\,\Lambda^2\;\leq\;\Lambda
\;,
\end{equation}
for $E>3\Lambda$ which will be assumed from now on. Actually, later on we will suppose that $E$ is even larger. Now writing out the M\"obius action and expressing $U^E_{N-1}$ in these formulas by $R^E_{N-1}$ according to \eqref{eq-RUrel} we obtain
\begin{align}
U^E_N
& 
\;=\;
\left[ ( \one - w_1 ) T^{-1} U^E_{N-1} T +     \one - w_2 \right]
\left[ (\one - w_3 ) T^{-1}  U^E_{N-1} T  +     \one - w_4 \right]^{-1}
\label{eq-UForm}
\\
& 
\;=\;
\left[\one - \tfrac{1}{2} w_1 - \tfrac{1}{2} w_2 -\ii (\one-w_1)E^{-1} + \tfrac{1}{2}( \one - w_1 ) T^{-1} R^E_{N-1} T  \right]
\,D_N^E
\nonumber
\\
& 
\;=\;
\left[\one -  (V+2\ii -\ii w_1)E^{-1} + \tfrac{1}{2}( \one - w_1 ) T^{-1} R^E_{N-1} T  \right]
\,D_N^E
\;,
\nonumber
\end{align}
where for further analysis we set
\begin{align*}
D_N^E
& 
\;=\;
2\,\left[ (\one - w_3 ) T^{-1}  U^E_{N-1} T  +     \one - w_4 \right]^{-1}
\\
& 
\;=\;
\left[\one - \tfrac{1}{2} w_3 - \tfrac{1}{2} w_4 -\ii (\one-w_3)E^{-1} + \tfrac{1}{2}( \one - w_3 ) T^{-1} R^E_{N-1} T  \right]^{-1}
\;.
\end{align*}
Let us also introduce $B^E_N$ by 
\begin{equation}
\label{eq-DId}
D_N^E
\;=\;
(\one-B^E_N)^{-1} 
\;=\;
\one +B_N^E D_N^E
\;,
\end{equation}
namely
\begin{align}
B_N^E
& 
\;=\;
\tfrac{1}{2} w_3 + \tfrac{1}{2} w_4 +\ii (\one-w_3)E^{-1} - \tfrac{1}{2}( \one - w_3 ) T^{-1} R^E_{N-1} T
\nonumber
\\
&
\;=\;
E^{-1}(V-\ii w_3 ) - \tfrac{1}{2} (1 - w_3 )  T^{-1} R_{N-1}^E T
\;.
\label{eq-BForm}
\end{align}
Next we need to express $R_N^E$ in terms of $R^E_{N-1}$. With $D_N^E =\one +B_N^E D_N^E$, one first finds
\begin{align*}
R^E_N
\;=\;
&
\left[
\one  - 2 \ii E^{-1}\one + \ii E^{-1} w_1  - E^{-1} V  +(1 - w_1 ) \tfrac{1}{2}  T^{-1} R_{N-1}^E T 
\right] [\one +B_N^E D_N^E] - (\one\,-\,2\,\ii\,E^{-1}) 
\\
\;=\;& 
\left[\ii E^{-1} w_1  - E^{-1} V  +(1 - w_1 ) \tfrac{1}{2}  T^{-1} R_{N-1}^E T 
\right] +  \\
& +  \left[
\one  - 2 \ii E^{-1}\one + \ii E^{-1} w_1  - E^{-1} V  +(1 - w_1 ) \tfrac{1}{2}  T^{-1} R_{N-1}^E T 
\right] B_N^E D_N^E
\\
\;=\;& 
\left[\ii E^{-1} w_1  - E^{-1} V  +(1 - w_1 ) \tfrac{1}{2}  T^{-1} R_{N-1}^E T 
\right] +  \\
& +  \left[
\one  - 2 \ii E^{-1}\one + \ii E^{-1} w_1  - E^{-1} V  +(1 - w_1 ) \tfrac{1}{2}  T^{-1} R_{N-1}^E T 
\right] B_N^E [\one +B_N^E D_N^E]
\\
\;=\;& 
\left[\ii E^{-1} w_1  - E^{-1} V  +(1 - w_1 ) \tfrac{1}{2}  T^{-1} R_{N-1}^E T 
\right] +  B_N^E + B_N^E B_N^E D_N^E\\
& +  \left[
- 2 \ii E^{-1}\one + \ii E^{-1} w_1  - E^{-1} V  +(1 - w_1 ) \tfrac{1}{2}  T^{-1} R_{N-1}^E T 
\right] B_N^E D_N^E
\;,
\end{align*} 
where in the last equality the definition \eqref{eq-DId} was used. Now regroup $B_N^E-E^{-1}V$ using again the definition of $B_N^E$, and use $w_1-w_3=2\ii E^{-1}$:
\begin{align}
R^E_N
\;=\;& 
\left[\ii E^{-1} w_1  +(1 - w_1 ) \tfrac{1}{2}  T^{-1} R_{N-1}^E T 
\right] + [-\ii E^{-1} w_3  - (1 - w_3 ) \tfrac{1}{2} T^{-1} R_{N-1}^E T] + B_N^E B_N^E D_N^E
\nonumber
\\
& +  \left[
- 2 \ii E^{-1}\one + \ii E^{-1} w_1  - E^{-1} V  +(1 - w_1 ) \tfrac{1}{2}  T^{-1} R_{N-1}^E T 
\right] B_N^E  D_N^E
\nonumber
\\
\;=\;& 
\left[\ii E^{-1} w_1  - w_1  \tfrac{1}{2}  T^{-1} R_{N-1}^E T 
\right] + [-\ii E^{-1} w_3  + w_3  \tfrac{1}{2} T^{-1} R_{N-1}^E T] + B_N^E B_N^E D_N^E
\nonumber
\\
& +  \left[
- 2 \ii E^{-1}\one + \ii E^{-1} w_1  - E^{-1} V  +(1 - w_1 ) \tfrac{1}{2}  T^{-1} R_{N-1}^E T 
\right] B_N^E  D_N^E
\nonumber
\\
\;=\;& 
-2 E^{-2}  - \ii E^{-1}  T^{-1} R_{N-1}^E T 
 + B_N^E B_N^E D_N^E
\nonumber
\\
& +  \left[
- 2 \ii E^{-1}\one + \ii E^{-1} w_1  - E^{-1} V  +(1 - w_1 ) \tfrac{1}{2}  T^{-1} R_{N-1}^E T 
\right] B_N^E  D_N^E
\;.
\label{eq-Riter}
\end{align}

We now proceed with the proof of the main estimate by estimating $B^E_N$ from \eqref{eq-BForm} using \eqref{eq-wEst}:
\begin{equation}
\label{eq-BEst}
\|B^E_N\|
\;\leq\;
E^{-1}
2\Lambda+\Lambda^3\|R^E_{N-1}\|
\;.
\end{equation}
Furthermore, one clearly has
\begin{equation}
\label{eq-DEst}
\|B^E_N\|
\;\leq\;\frac{1}{2}
\qquad\Longrightarrow\qquad
\|D^E_N\|\;\leq\;2
\;.
\end{equation}
Below we will always assure to be in this case. Now let us estimate $R^E_N$ using \eqref{eq-Riter}:
\begin{align}
\|R^E_N\|
\;\leq\;
& 2 E^{-2}
\,+\,
E^{-1}\Lambda^2\|R^E_{N-1}\|
\,+\,
\|B^E_N\|^2\,\|D^E_N\|
\nonumber
\\
& \,+\,
\left[
E^{-1}(2+2\Lambda)   +(1 +\Lambda) \tfrac{1}{2} \Lambda^2\| R_{N-1}^E\| 
\right]\| B_N^E \|\,\| D_N^E\|
\nonumber
\displaybreak[0]\\ 
\;\leq\;
&
2 E^{-2}
\,+\,
\big[\|B^E_N\|\,+\,4E^{-1}\Lambda \big]\|B^E_N\|\,\|D^E_N\|
\,+\,
\big[E^{-1}\Lambda^2\,+\,\Lambda^3 \| B_N^E \|\,\| D_N^E\|\big] \|R^E_{N-1}\|
\nonumber
\displaybreak[0]\\ 
\;\leq\;
&
2 E^{-2}
\,+\,
6E^{-1}\Lambda \|B^E_N\|\,\|D^E_N\|
\,+\,
\big[E^{-1}\Lambda^2\,+\,2\Lambda^3 \| B_N^E \|\,\| D_N^E\|\big] \|R^E_{N-1}\|
\nonumber
\displaybreak[0]\\ 
\;\leq\;
&
2 E^{-2}
\,+\,
12E^{-2}\Lambda^2 \,\|D^E_N\|
\,+\,
\big[6E^{-1}\Lambda^4\|D^E_N\|\,+\,E^{-1}\Lambda^2\,+\,2\Lambda^3 \| B_N^E \|\,\| D_N^E\|\big] \|R^E_{N-1}\|
\nonumber
\displaybreak[0]\\ 
\;\leq\;
&
E^{-2}(2
\,+\,
24\, \Lambda^2)
\,+\,
\big[13\,E^{-1}\Lambda^4\,+\,4\,\Lambda^3 \| B_N^E \|\,\big] \|R^E_{N-1}\|
\;,
\label{eq-REst}
\end{align}
where in the last steps \eqref{eq-BEst} and $\| D_N^E\|\leq 2$ were used. The r.h.s. is still linear in $\|R^E_{N-1}\|$, but we can now use \eqref{eq-BEst} again to obtain an estimate containing a quadratic term:
\begin{align}
\|R^E_N\|
\;\leq\;
E^{-2}
\,26 \Lambda^2
\,+\,
21\,E^{-1}\Lambda^4\,\|R^E_{N-1}\|
\;+\;
4\,\Lambda^6\, \|R^E_{N-1}\|^2
\;.
\label{eq-REst3}
\end{align}
Before going on, let us estimate the initial conditions for the two equations \eqref{eq-BEst} and \eqref{eq-DEst}. From
$$
U_1^E 
\;=\; 
[ (E \one - V_1) - \ii \one ] [(E \one - V_1) + \ii \one ]^{-1} 
\;=\; 
[ \one - V_1 E^{-1} - \ii E^{-1} \one ] 	[  \one - E^{-1}  (V_1 - \ii \one) ]^{-1}
\;,
$$
one deduces
$$
R^E_1
\;=\;
E^{-2}(-2-2\,\ii\,V_1)(\one-E^{-1}V_1+\ii\,E^{-1})^{-1}
\;.
$$
As $\|V_1\| \leq \Lambda$ the inverse can be estimated by $2$  for $E >4\Lambda> 2(\Lambda+1)$, and thus follows the initial estimate
$$
\|R_1^E\|
\;\leq\; 
E^{-2} (2+2\Lambda)2
\;\leq\;
8 E^{-2}\Lambda
\;.
$$
Due to \eqref{eq-BEst}, this shows
$$
\|B^E_2\|
\;\leq\;
E^{-1}
2\Lambda+8 E^{-2} \Lambda^4
\;,
$$
so for $E$ large, in particular, $\|B^E_2\|\leq\frac{1}{2}$ as required in \eqref{eq-DEst}. For the final contraction argument, let us set
$$
r_N
\;=\;
E^2\,\|R^E_N\|
\;,
$$
and introduce the function $f_E:\RM_\geq\to\RM_\geq$ by
$$
f_E(x)
\;=\;
26\, \Lambda^2
\,+\,
21\,E^{-1}\Lambda^4\,x
\;+\;
4\,E^{-2}\,\Lambda^6\, x^2
\;.
$$
Then, because \eqref{eq-DEst} holds, \eqref{eq-REst3} now reads
$$
r_N\;\leq\;f_E(r_{N-1})
\;.
$$
If $E$ is sufficiently large, then $f_E$ has an attractive fixed point $x_E$ with a basin of attraction $[0,b_E)$ where $b_E$
is the second fixed point, a repeller. One has $b_E\to\infty$ for $E\to\infty$. Moreover, $x_E\leq c\, \Lambda^2$ for some constant $c\geq 8$ and $E$ sufficiently large. Furthermore, possibly for $E$ even larger, $b_E>c\,\Lambda^2$. Since $r_1\leq 8\Lambda\leq c\,\Lambda^2$, then $r_N\leq c\,\Lambda^2$ for all $N\geq 1$. This implies the result. 
\hfill $\Box$

\vspace{.2cm}

The proof of Proposition~\ref{prop-PrueferBound} implies that $\partial_E U_N^E$ grows at most polynomially in $E$. In the following, we will shows that it actually decrease for large $E$. In principle, one could be as explicit about the constants in the following, but we refrained from doing so.

\begin{lemma} 
\label{lem-Deribound}
There is a constant $C'$ depending on $\Lambda$ and an $E_0$ such that for all $N\geq 1$ and  $|E| > E_0$: 
$$
\|\partial_E U_N^E - 2 \ii  E^{-2}\one \|
\;\leq \;
\frac{C' }{|E|^3}
\;.
$$
\end{lemma}

\noindent{\bf Proof.} We will use several notations from the proof of Proposition~\ref{prop-asymptotics}, in particular $w_j$, $D^E_N$ and $B^E_N$. Furthermore, $T=T_N$ and $V=V_N$ will not carry an index. Let us also introduce a notation for the term that needs to be estimated:
$$
P_N^E \;=\; 
\partial_E U_N^E - 2 \ii  E^{-2}\one
\;.
$$
For the calculation of the derivative $\partial_E U_N^E $, let us start from \eqref{eq-UForm}:
\begin{align*}
P_N^E
\,=\,
& 
\tfrac{1}{2} \left[ 
- (\partial_E w_1) T^{-1} U^E_{N-1} T + ( \one - w_1 ) T^{-1} \partial_E U^E_{N-1} T - \partial_E w_2
\right]
D^E_N
\\
& -\, \tfrac{1}{2}
\left[ ( \one - w_1 ) T^{-1} U^E_{N-1} T +     \one - w_2 \right]D^E_N 
\\
&
\;\;\;\;\;
\cdot\, \tfrac{1}{2}\left[ 
 - (\partial_E w_3) T^{-1} U^E_{N-1} T + ( \one - w_3 ) T^{-1} \partial_E U^E_{N-1} T - \partial_E w_4
 \right]
D^E_N \,-\, 2\ii  E^{-2}\one
\;. 
\end{align*}
Now let us replace $\partial_E U_{N-1}^E =P_{N-1}^E + 2 \ii  E^{-2}\one$ and split $U_{N-1}^E=(U_{N-1}^E-\one)+\one$:
\begin{align*}
P_N^E
\,=\,
& 
\tfrac{1}{2} \left[ 
- (\partial_E w_1) T^{-1} (U^E_{N-1}-\one) T + ( \one - w_1 ) (T^{-1} P^E_{N-1} T +2 \ii  E^{-2} \one)- \partial_E (w_1+w_2)
\right]
D^E_N
\\
& -\, \tfrac{1}{2}
\left[ ( \one - w_1 ) T^{-1} (U^E_{N-1}-\one) T +    2\, \one -(w_1+ w_2) \right]D^E_N 
\\
&
\;\;\;\;\;
\cdot\, \tfrac{1}{2}\left[ 
 - (\partial_E w_3) T^{-1} (U^E_{N-1}-\one) T + ( \one - w_3 )( T^{-1} P^E_{N-1} T +2\ii E^{-2}\one)- \partial_E (w_3+w_4)
 \right]
D^E_N 
\\
& \;\;\;\;\;
\;\;\;\;\;
\,-\, 2\ii  E^{-2}\one
\;. 
\end{align*}
Now from the definitions
$$
w_1+w_2\;=\;2E^{-1}(V+\ii\one)\;,
\qquad
w_3+w_4\;=\;2E^{-1}(V-\ii\one)\;,
$$
so that
$$
\partial_E(w_1+w_2)\;=\;-2E^{-2}(V+\ii\one)\;,
\qquad
\partial_E(w_3+w_4)\;=\;-2E^{-2}(V-\ii\one)\;.
$$
Let us replace these latter two equations in the expression for $P_N^E$:
\begin{align*}
P_N^E
\,=\,
& 
\tfrac{1}{2} \left[ 
- (\partial_E w_1) T^{-1} (U^E_{N-1}-\one) T + ( \one - w_1 ) T^{-1} P^E_{N-1} T -2 \ii  E^{-2} w_1+2E^{-2}(V+2\ii\one)
\right]
D^E_N
\\
& -\, 
\left[\one+\tfrac{1}{2} ( \one - w_1 ) T^{-1} (U^E_{N-1}-\one) T -\tfrac{1}{2}(w_1+ w_2) \right]D^E_N 
\\
&
\;\;\;\;\;
\cdot\, \tfrac{1}{2}\left[ 
 - (\partial_E w_3) T^{-1} (U^E_{N-1}-\one) T + ( \one - w_3 ) T^{-1} P^E_{N-1} T -2\ii E^{-2}w_3 +2E^{-2}V
 \right]
D^E_N 
\\
& \;\;\;\;\;
\;\;\;\;\;
\,-\, 2\ii  E^{-2}\one
\;. 
\end{align*}
To see that the terms of order $E^{-2}$ indeed cancel out, we can now replace  $D^E_N=\one+B^E_ND^E_N$ as given in \eqref{eq-DId} several times:
\begin{align*}
P_N^E
\,=\,
& 
\tfrac{1}{2} \left[ 
- (\partial_E w_1) T^{-1} (U^E_{N-1}-\one) T + ( \one - w_1 ) T^{-1} P^E_{N-1} T -2 \ii  E^{-2} w_1\right]
\\
& 
+\,\tfrac{1}{2} \left[ 
- (\partial_E w_1) T^{-1} (U^E_{N-1}-\one) T + ( \one - w_1 ) T^{-1} P^E_{N-1} T -2 \ii  E^{-2} w_1+2E^{-2}(V+2\ii\one)
\right]
B^E_ND^E_N
\\
& -\, 
\left[\one+\tfrac{1}{2} ( \one - w_1 ) T^{-1} (U^E_{N-1}-\one) T -\tfrac{1}{2}(w_1+ w_2) \right]B^E_ND^E_N 
\\
&
\;\;\;\;\;
\cdot\, \tfrac{1}{2}\left[ 
 - (\partial_E w_3) T^{-1} (U^E_{N-1}-\one) T + ( \one - w_3 ) T^{-1} P^E_{N-1} T -2\ii E^{-2}w_3 +2E^{-2}V
 \right]
D^E_N 
\\
& -\, 
\left[\one+\tfrac{1}{2} ( \one - w_1 ) T^{-1} (U^E_{N-1}-\one) T -\tfrac{1}{2}(w_1+ w_2) \right]
\\
&
\;\;\;\;\;
\cdot\, \tfrac{1}{2}\left[ 
 - (\partial_E w_3) T^{-1} (U^E_{N-1}-\one) T + ( \one - w_3 ) T^{-1} P^E_{N-1} T -2\ii E^{-2}w_3 +2E^{-2}V
 \right]
B^E_ND^E_N 
\\
& -\, 
\left[\one+\tfrac{1}{2} ( \one - w_1 ) T^{-1} (U^E_{N-1}-\one) T -\tfrac{1}{2}(w_1+ w_2) \right]
\\
&
\;\;\;\;\;
\cdot\, \tfrac{1}{2}\left[ 
 - (\partial_E w_3) T^{-1} (U^E_{N-1}-\one) T + ( \one - w_3 ) T^{-1} P^E_{N-1} T -2\ii E^{-2}w_3
 \right]
\\
& -\, 
\left[\tfrac{1}{2} ( \one - w_1 ) T^{-1} (U^E_{N-1}-\one) T -\tfrac{1}{2}(w_1+ w_2) \right]E^{-2}V
\;. 
\end{align*}
The only terms which still have to be canceled out are $\frac{1}{2}T^{-1}P_{N-1}^ET$ appearing in the first line and the second to last summand:
\begin{align*}
P_N^E
\,=\,
& 
\tfrac{1}{2} \left[ 
- (\partial_E w_1) T^{-1} (U^E_{N-1}-\one) T - w_1  T^{-1} P^E_{N-1} T -2 \ii  E^{-2} w_1\right]
\\
& 
+\,\tfrac{1}{2} \left[ 
- (\partial_E w_1) T^{-1} (U^E_{N-1}-\one) T + ( \one - w_1 ) T^{-1} P^E_{N-1} T -2 \ii  E^{-2} w_1+2E^{-2}(V+2\ii\one)
\right]
B^E_ND^E_N
\\
& -\, 
\left[\one+\tfrac{1}{2} ( \one - w_1 ) T^{-1} (U^E_{N-1}-\one) T -\tfrac{1}{2}(w_1+ w_2) \right]B^E_ND^E_N 
\\
&
\;\;\;\;\;
\cdot\, \tfrac{1}{2}\left[ 
 - (\partial_E w_3) T^{-1} (U^E_{N-1}-\one) T + ( \one - w_3 ) T^{-1} P^E_{N-1} T -2\ii E^{-2}w_3 +2E^{-2}V
 \right]
D^E_N 
\\
& -\, 
\left[\one+\tfrac{1}{2} ( \one - w_1 ) T^{-1} (U^E_{N-1}-\one) T -\tfrac{1}{2}(w_1+ w_2) \right]
\\
&
\;\;\;\;\;
\cdot\, \tfrac{1}{2}\left[ 
 - (\partial_E w_3) T^{-1} (U^E_{N-1}-\one) T + ( \one - w_3 ) T^{-1} P^E_{N-1} T -2\ii E^{-2}w_3 +2E^{-2}V
 \right]
B^E_ND^E_N 
\\
& -\, 
\left[\one+\tfrac{1}{2} ( \one - w_1 ) T^{-1} (U^E_{N-1}-\one) T -\tfrac{1}{2}(w_1+ w_2) \right]
\\
&
\;\;\;\;\;
\cdot\, \tfrac{1}{2}\left[ 
 - (\partial_E w_3) T^{-1} (U^E_{N-1}-\one) T  - w_3  T^{-1} P^E_{N-1} T -2\ii E^{-2}w_3
 \right]
\\
& -\, 
\left[\tfrac{1}{2} ( \one - w_1 ) T^{-1} (U^E_{N-1}-\one) T -\tfrac{1}{2}(w_1+ w_2) \right]
\left[\tfrac{1}{2}\, T^{-1} P^E_{N-1} T + E^{-2}V\right]
\;. 
\end{align*}
For the following estimates, we now use $\|U_N^E-\one\|\leq cE^{-1}$ as follows from Proposition~\ref{prop-asymptotics}, for some constant $c$ (which takes increasing values in the following).  Furthermore, $\|w_j\|\leq c E^{-1}$ and $\|\partial_Ew_j\|\leq cE^{-2}$ as well as $\|B^E_N\|\leq cE^{-1}$ and $\|D^E_N\|\leq 2$, see the proof of Proposition~\ref{prop-asymptotics}. Carefully checking all terms, this leads to
$$
\|P_N^E\|
\;\leq\;
c\,E^{-3}\,+c\,E^{-1}\|P_{N-1}^E\|
\;.
$$
Setting now $p^E_N=E^3\|P_N^E\|$, it follows that $p^E_N\leq c+cE^{-1}p_{N-1}^E\leq c(1-cE^{-1})^{-1}$. This concludes the proof.
\hfill $\Box$

\begin{proposi}
\label{prop-asymptotics2}
There is a constant $C''$ and an $E_0$ such that for all $N\geq 1$ and  $|E| > E_0$: 
$$
\|S_N^E \,-\, 2N^{-1}  E^{-2} \one \| \;\leq\;\frac{C''}{N|E|^3} 
\;.
$$
\end{proposi}

\noindent {\bf Proof.} Let us begin by splitting the contributions as follows:
\begin{align*}
N\;\|S_N^E \,-\, 2N^{-1}  E^{-2} \one \| 
& 
\;\leq\;
\|(U_N^E)^* \partial_E U_N^E - 2 \ii  (U_N^E)^* E^{-2}   \|
\; +\;
\|2 \ii  (U_N^E)^* E^{-2} - 2 \ii   E^{-2} \one\|
\\
& 
\;=\;
\| \partial_E U_N^E - 2 \ii  E^{-2}\one \|
\; +\;
2E^{-2}\,\|(U_N^E)^* -  \one \|
\;.
\end{align*}
Now the result follows from Proposition~\ref{prop-asymptotics} and Lemma~\ref{lem-Deribound}.
\hfill $\Box$

\section{Finite volume approximations}
\label{sec-FiniteApp}

The strategy for the proof of Theorem~\ref{theo-osci} is to control the finite volume approximations of both $\Nn_N$ and $S^E_N$ in the vertical, infinite direction. Such approximations follow from standard ergodic properties of the operator entries of the Jacobi operator \eqref{eq-matrix}. On the other hand, for the approximating Jacobi matrix with matrix entries (of finite dimension) Sturm-Liouville-type oscillation theory as developed in \cite{SB} will be exploited below. 

\vspace{.2cm}

To construct the finite volume approximants, let 
$$
\pi_L\;:\; \ell^2(\ZM^d) \,\to \,\CM^{(2L+1)^d}\;\cong\;\ell^2(\{-L,\ldots,L\}^d)\;\subset\; \ell^2(\ZM^d)
$$ 
be the surjective partial isometry given by the restriction of the sequences. Then the restrictions  of $T_n$ and $V_n$ with Dirichlet boundary condition are $\pi_L T_n\pi_L^*$ and $\pi_L V_n\pi_L^*$. Up to boundary terms localized close to the boundary of the cube $\{-L,\ldots,L\}^d\subset \ZM^d$, these restrictions are the approximants $T_{n,L}$ and $V_{n,L}$ to be used in the following. We will suppose that the boundary terms can be chosen such that for some $L_0$
\begin{equation}
\label{eq-HypFiniteBound}
\Lambda_L \;=\;
\sup_{n\geq 1}\;\{\|T_{n,L}\|,\|T^{-1}_{n,L}\|,\|V_{n,L}\|\}
\;\leq\;2\,\Lambda\;<\;\infty
\;,
\qquad
\mbox{\rm for } 
L\geq L_0
\;.
\end{equation}
This excludes topological systems with protected boundary states which can possibly destroy the invertibility of $T_{n,L}$, but is satisfied for the restrictions in the standard Anderson model discussed in the introduction. Once the approximants $T_{n,L}$ and $V_{n,L}$ are chosen such that \eqref{eq-HypFiniteBound}, one can define the finite volume approximation of the Hamiltonian:
\begin{equation}
\label{eq-matrix2}
H_{N,L}
\;=\;
\left(
\begin{array}{ccccccc}
V_{1  ,L}     & T_{2,L}  &        &        &         &    \\
T_{2,L}^*      & V_{2 ,L}   &  T_{3,L}  &        &         &        \\
            & T_{3,L}^* & V_{3,L}    & \ddots &         &        \\
            &        & \ddots & \ddots & \ddots  &        \\
            &        &        & \ddots & V_{N-1,L} & T_{N,L}   \\
    &        &        &        & T_{N,L}^*  & V_{N,L}
\end{array}
\right)
\;.
\end{equation}
It is a selfadjoint matrix of size $N(2L+1)^d$. As such, it has a normalized eigenvalue counting function
$$
\Nn_{N,L}(E)
\;=\;
\frac{1}{N(2L+1)^d}\;\Tr\big(\chi(H_{N,L}\leq E)\big)
\;.
$$
The function $E\in\RM\mapsto \Nn_{N,L}(E)$ is increasing with asymptotics $\Nn_{N,L}(-\infty)=0$ as well as $\Nn_{N,L}(\infty)=1$. For fixed $N$, the following result is a standard consequence of the Birkhoff's ergodic theorem, {\it e.g.} \cite{Bel,PF}.

\begin{proposi}
\label{prop-DOSconv}
Suppose that the $T_n$'s and $V_n$'s are covariant and local operators on $\ell^2(\ZM^d)$ with approximants $T_{n,L}$ and $V_{n,L}$ as described above. For any fixed $N\in\NM$ and $E\in\RM$, one has $\PM$-almost surely
$$
\lim_{L\to\infty} \Nn_{N,L}(E)
\;=\;
\Nn_N(E)
\;.
$$
The convergence also holds in expectation. 
\end{proposi}

Uniform estimates in energy on the convergence of the integrated density of states have recently been obtained in \cite{SSV}, but this will not be needed here. As $\Nn_N(E)$ is one of the quantities in the main result, Theorem~\ref{theo-osci}, we will now calculate $\Nn_{N,L}(E)$ by means of matrix-valued oscillation theory. For this purpose, we first need to introduce for any finite $L$ the transfer matrices $\trans_{n,L}^E$, the Pr\"ufer phases $U^E_{N.L}$ and their derivative. For any $E\in\RM$ and with the convention $T_{1,L} = \one$, we set
\begin{equation}
\label{eq-TM-finite}
\trans_{n,L}^E
\;=\;
\left(\begin{array}{cc}(E\,{\bf 1}\,-\,V_{n,L})\,T_{n,L}^{-1} & - T_{n,L}^* \\
T_{n,L}^{-1} & 0
\end{array}
\right)
\;,
\qquad
n=1,\ldots,N
\;.
\end{equation}
This matrix is again an $\Ii$-unitary defined as in Section~\ref{sec-Krein}, albeit with $\Ii$ as in \eqref{eq-JIdef} with matrix entries of size $(2L+1)^d$. With a Cayley transform $\Cc$ also defined as in \eqref{eq-Cayley} with matrix entries, the matrix $\Cc\trans_{n,L}^E\Cc^*$ is $\Jj$-unitary and hence acts on the $\Jj$-Lagrangian subspaces of the finite dimensional Krein space $(\CM^{(2L+1)^d}\otimes\CM^2,\Jj)$, with $\Jj$ again defined as in \eqref{eq-JIdef} with matrix entries of size $(2L+1)^d$. This action is again implemented as the M\"obius action on the unitary group of the same size. Hence one has,  {\it cf.} \eqref{eq-MoebRep},
\begin{equation}
\label{eq-Pruefer-finite}
U^E_{N,L}\;=\;
\Cc\,\trans_{N,L}^E\cdots\trans_{1,L}^E \,\Cc^*
\cdot
\one
\;.
\end{equation}
Finally let us introduce the phase velocity matrices by
\begin{equation}
\label{eq-PrueferDeri-finite}
S^E_{N,L}
\; =\; \frac{1}{\ii {N}} \big(U^E_{N,L}\big)^* \partial_E U^E_{N,L}
\;.
\end{equation}
Note that this definition differs from that found in \cite{SB} by the factor $N^{-1}$. By the same proof as that of Proposition~\ref{prop-derivbound} one has $S^E_{N,L}\geq 0$. Furthermore, for these finite volume approximations, the bounds in Propositions~\ref{prop-PrueferBound}, \ref{prop-asymptotics} and \ref{prop-asymptotics2} all hold because they were merely using norm estimates based on the standing hypothesis. We omit a detailed proof of the following fact.

\begin{proposi}
\label{prop-FiniteAppBounds}
Let $L\geq L_0$ so that \eqref{eq-HypFiniteBound} holds. Then all the bounds in Propositions~\ref{prop-PrueferBound}, \ref{prop-asymptotics} and \ref{prop-asymptotics2} hold for $U^E_{N,L}$ and $S^E_{N,L}$ instead of $U^E_N$ and $S^E_N$. 
\end{proposi}

Propostion~\ref{prop-FiniteAppBounds} provides pointwise estimates not requiring any ergodicity properties. These latter are used again for the next result which is similar to Proposition~\ref{prop-DOSconv}. As the techniques of proof are standard, we do not provide lengthy details.

\begin{proposi}
\label{prop-ConvS}
Under the same hypothesis as in Proposition~\ref{prop-DOSconv} and provided that \eqref{eq-HypFiniteBound} holds, one has for any fixed $E$ and $N$ and with $\PM$-almost sure convergence
$$
\lim_{L\to\infty} \;
\frac{1}{(2L+1)^d}\;\Tr(S^E_{N,L})
\;=\;
\Tt(S^E_N)
\;.
$$
The convergence also holds in expectation. 
\end{proposi}

\noindent {\bf Sketch of proof.} By construction, the matrix entries $T_{n,L}$ and $V_{n,L}$ of $\Mm^E_{n,L}$ as given in \eqref{eq-TM-finite} are finite volume restrictions of covariant operators $T_n$ and $V_n$, up to boundary terms. Because these operators are also local (namely in the C$\mbox{*}$\hbox{-}algebra $\Aa$ described in Section~\ref{sec-covariant}), the geometric resolvent identity shows that also $(T_{n,L})^{-1}$ is the finite volume restriction of $(T_n)^{-1}$ up to boundary terms (compact, but not of finite range any more). As this property is inherited to products of such operators, also the matrix entries of $\Mm^E_{N,L}\cdots  \Mm^E_{1,L}$ are finite volume restrictions of the matrix entries of $\Mm^E_{N}\cdots  \Mm^E_{1}$, up to boundary terms. In conclusion, also $U^E_{N,L}$ given in \eqref{eq-Pruefer-finite} is a finite volume restriction of $U^E_{N}$ given in \eqref{eq-MoebRep}, up to boundary terms. As the Pr\"ufer phases are rational in $E$, also $S^E_{N,L}$ is a finite volume restriction of $S^E_{N}$ up to boundary terms. Therefore Birkhoff's theorem implies as in \cite{Bel,PF} the stated almost sure convergence.
\hfill $\Box$

\vspace{.2cm}

The final result in the section is one of the main results of oscillation theory for matrix valued Jacobi matrices as developed in \cite{SB,SB2}, see also \cite{DK,KHZ} and \cite{JONNF}. We do not provide any proof, as the statement is precisely the estimate given in equation~(40) of \cite{SB}.

\begin{theo}
\label{theo-MaslovEst}
For any Jacobi matrix $H_{N,L}$ with matrix entries, the eigenvalue counting function $\Nn_{N,L}$ is approximated by the speed matrix $S_{N,L}$ uniformly in $L$:
\begin{equation}
\label{eq-Bound40}
\left|\,
\Nn_{N,L}(E)
\;-\;\frac{1}{2\pi}\,\int^E_{-\infty}de\;\frac{1}{(2L+1)^d}\;\Tr(S^e_{N,L})
\,\right|
\;\leq\;\frac{2}{N}
\;.
\end{equation}
\end{theo}

\vspace{.3cm}

\section{Proof of the main result}
\label{sec-Proof}

In this section, we resemble the estimates and facts stated above to provide a proof of the main result of the paper.

\vspace{.2cm}

\noindent {\bf Proof } of Theorem~\ref{theo-osci}. Let us begin by estimating the expression in Theorem~\ref{theo-osci} by a sum of three terms, namely with $\EE$ denoting the average w.r.t. $\PM$ and for $L$ to be chosen later
\begin{align}
\left|
\,\Nn_N(E)\,-\,
\frac{1}{2\pi}\,\int^E_{-\infty} de\;\Tt(S^e_N)\,
\right|
\;\leq\;
&
\left|\,
\Nn_N(E)\,-\,\EE\;\Nn_{N,L}(E)
\,\right|
\nonumber
\\
& \;+\;
\EE\;
\left|\,\,
\Nn_{N,L}(E)
\;-\;\frac{1}{2\pi}\,\int^E_{-\infty}de\;\frac{1}{(2L+1)^d}\,\Tr(S^e_{N,L})
\,\right|
\label{eq-Est2}
\\
& \;+\;
\frac{1}{2\pi}\,\left|
\,\int^E_{-\infty}de\;\EE\;
\frac{1}{(2L+1)^d}\,\Tr(S^e_{N,L})
\,-\,\int^E_{-\infty} de\;\Tt(S^e_N)
\,
\right|
\;.
\nonumber
\end{align}
The first contribution on the r.h.s. of \eqref{eq-Est2} can be made arbitrarily small, say less than $N^{-1}$, by choosing $L$ sufficiently large, see Proposition~\ref{prop-DOSconv}. The second summand in \eqref{eq-Est2} is less than or equal to $2N^{-1}$ for any $L$ by Theorem~\ref{theo-MaslovEst}. The last summand will again be split in several contributions, namely for large $E_0>0$ as in Section~\ref{sec-Asymp} we  bound it as follows:
\begin{align}
\int^E_{-\infty}de\;\left|
\,\EE\;\frac{1}{(2L+1)^d}\,\Tr(S^e_{N,L})
\,-\,\Tt(S^e_N)\,
\right|
\;\leq\;
&
\int^{-E_0}_{-\infty} de\;\Tt(S^e_N)
\nonumber 
\\
& +\;
\int^{-E_0}_{-\infty} de\;\EE\;
\frac{1}{(2L+1)^d}\;\Tr(S^e_{N,L})
\label{eq-Est3}
\\
& +\;
\int^E_{-E_0}de\;\left|
\,\EE\;\frac{1}{(2L+1)^d}\,\Tr(S^e_{N,L})
\,-\,\Tt(S^e_N)
\,
\right|
\nonumber
\;.
\end{align}
From the estimate $|\Tt(A)|\leq \|A\|$ combined with Proposition~\ref{prop-asymptotics2} it follows that
\begin{equation}
\label{eq-Est4}
\int^{-E_0}_{-\infty} de\;\Tt(S^e_N)
\;\leq\;
\frac{C}{NE_0}
\;,
\end{equation}
 for some constant $C$. Using Proposition~\ref{prop-FiniteAppBounds} one concludes that the same estimate holds for the second summand in \eqref{eq-Est3}. Let us note that the bound \eqref{eq-Est4} also proves the theorem for $E<-E_0$, and similarly for $E>E_0$.  As to the last summand in \eqref{eq-Est3}, Proposition~\ref{prop-ConvS} assures the pointwise convergence of the integrand for each $e$. Moreover, Proposition~\ref{prop-FiniteAppBounds} states $\sup_{L>L_0} \Big( \sup_{e \in [-E_0,E]}\| \partial_e U_{N,L}^e \| \Big)\leq K$, which combined with the Arzel\`{a}-Ascoli theorem  shows that the convergence of a subsequence of the integrand is uniform on the compact set $[-E_0,E]$.  Therefore, also the last summand in \eqref{eq-Est3} converges to $0$ as $L\to\infty$. In particular, it can be made smaller than $N^{-1}$ for $L$ sufficiently large. In conclusion, all terms on the r.h.s. of \eqref{eq-Est2} can be made smaller than a constant times $N^{-1}$ by choosing $L$ sufficiently large. This concludes the proof.
\hfill $\Box$

\vspace{.5cm}


\end{document}